\documentclass[
reprint,
superscriptaddress,
groupedaddress,
nofootinbib,
nobibnotes,
 amsmath,amssymb,
 aps,
prb,
]{revtex4-1}

\usepackage[normalem]{ulem}
\usepackage{siunitx}
\usepackage{graphicx}
\usepackage{dcolumn}
\usepackage{bm}
\usepackage{epstopdf}
\usepackage{bbm}
\usepackage{amsmath}
\usepackage{color}
\usepackage[hidelinks]{hyperref}
\usepackage{commath}
\usepackage{tabularx}
\usepackage{booktabs}

\usepackage{bbold}

\begin{document}

\preprint{APS/123-QED}

\title{ 
Particle-hole symmetry broken solutions in graphene nanoribbons: \\ a multi-orbital, mean-field perspective
}
\author{T. Schmirander}%
\email{tschmira@physnet.uni-hamburg.de}
\affiliation{%
	I. Institute for Theoretical Physics, Universit\"at  Hamburg 
D-22761 Hamburg, Germany
}
\author{D. Pfannkuche }%
\affiliation{%
	I. Institute for Theoretical Physics, Universit\"at  Hamburg 
D-22761 Hamburg, Germany
}
\author{M. Prada}%
\email{mprada@physnet.uni-hamburg.de}
\affiliation{%
	I. Institute for Theoretical Physics, Universit\"at  Hamburg 
D-22761 Hamburg, Germany
}
\date{\today}
\begin{abstract}
Mean-field theories have since long predicted edge magnetism in graphene nanoribbons, 
where the order parameter is given by the local magnetization. 
However, signatures of edge magnetism appears elusive in the experiments, suggesting another class of solutions. 
We employ a self-consistent mean field approximation within a multi-orbital tight-binding model 
and obtain particle-hole symmetry broken solutions, where the local filling plays the role of the order parameter. 
Unlike the magnetic edge solutions, 
these are topologically non-trivial and show zero local magnetization. 
A small and a large doping regime are studied, and a free energy minimum for finite hole doping is encountered,
which may serve as an explanation for the absence of experimental evidence for magnetic edge states in
zigzag graphene nanoribbons.
The electronic interaction 
may increase the finite \(d\)-orbital occupation, 
which leads to a change of the effective Coulomb interaction of the dominant \(p_z\)-orbitals. 
Our findings indicate that the non-magnetic solution for finite hole doping becomes energetically preferred, 
compared to the magnetic phases at half-filling, once thermal fluctuations or unintentional doping from the substrate are considered. 
This result persists even in the presence of the \(d\)-orbitals and the Coulomb interaction therein.
\end{abstract}
\keywords{Topological Insulators, Magnetism, Graphene}
\maketitle

\section{Introduction}
Graphene nanoribbons with zigzag edges belong to the class of topological matter with insulating bulk and conducting 
edges  \cite{Fruchart2013,Asboth2016, GosalbezMartines2012}.
The Fermi level crossing states were predicted to give rise to a novel type of
magnetic ordering \cite{Yazyev2011PRB}, leading to appealing carbon-based magnetic materials and 
stimulating a large number of proposals of novel graphene-based spintronic devices 
\cite{Yazyev2010, katsnelson2008,Richter2008,RojasPRL2009,Pascual2020,li2021topological}. 
Many experiments have ever since claimed to prove the existence of magnetism at the edges 
\cite{Kobayashi2005, Shibayama2000}, but the accuracy of these findings allow for explanations 
other than the occurrence of magnetic edge states \cite{Joly2010}. 
A gap opening in the local density of states has been found, which is consistent with magnetic ordering of the edge phases 
\cite{Li2014,Tao2011, Magda2014}, but experiments usually probe only either the edge character of electronic states in 
graphene or the alignment of the spin degree of freedom. This poses the first problem of experimental verification of the topologically insulating phase of graphene that is still unresolved: Is the ground state of graphene nanoribbons truly non-magnetic or has it just not been measured
yet? 

The description of non-interacting graphene is usually performed by employing the Kane-Mele model 
\cite{KaneMele2005a,KaneMele2005b,Goldman2012}, which distinguishes between a trivial and a 
non-trivial phase in terms of external parameters.  
This is commonly extended to the interacting Kane-Mele-Hubbard model \cite{Rachel2018,Zheng2011,Hohenadler2014,Cao2013}. 
While such considerations apply to bulk graphene, coupling of the edge states in terminated structures leads 
to an energy gap in the magnetic ground state \cite{Luo2020, Soriano2010,Lado2015}, which can be described 
by an \emph{inter-edge superexchange} \cite{Jung2009b,FernandesRossier2008}. 
This mechanism predicts a zero-spin ground state,  where the magnetic edges show opposite polarization.  
Different types of inter-edge magnetism can occur in graphene nanoribbons at half-filling \cite{GosalbezMartines2012,Wakabayashi1998}, 
namely, the anti-ferromagnetic (AFM) and  the ferromagnetic state (FM), 
but other states with different degrees of edge magnetism are also possible  once doping is considered 
\cite{Jung2009a, Yazyev2010}.
The particle-hole symmetry (PHS) is broken in a doped system, and a completely non-magnetic phase is possible, 
as the Fermi energy is located away from the magnetic instability. 
This phase may play a central role in realistic structures, as unintentional doping or 
 commonly used substrates (SiO\(_2\) \cite{Fan2012}, 
SiC \cite{varchon2007} or h-BN \cite{henckHBN})   
has been shown to shift the Fermi level off the half filling condition at the Dirac points (DPs), 
and hence, broken PHS is expected in the experiments. 
Edge states in graphene nanoribbons are commonly examined by using density-functional theory (DFT) of the 
\(\pi\)-orbitals in graphene \cite{Correa2018, Polini2008}, where the mean-field method remains a good approximation for 
realistic interactions \cite{Feldner2010}. 

The atomic or intrinsic \(d\)-orbital spin-orbit coupling 
becomes relevant for the size of the gap at the Dirac points of graphene \cite{Konschuh2010, jonas} and should not be neglected. 
Moreover, it provides helicity to the edge states \cite{KaneMele2005b, pradaPRB21b,pradaPRB21a}.  
This leads to the second problem of detecting edge magnetism in graphene, namely, the influence of the electronic 
interaction of the \(d\)-orbital electrons on the \(\pi\)-orbitals in a realistic --or doped-- system, 
which has not been examined to this date. 

This manuscript is intended to address these two problems by defining a multi-orbital tight-binding model with 
\(p_z\)- and \(d\)-orbitals, and is organized as follows:
In Section \ref{methods} we describe the multi-orbital tight-binding model and the self-consistent mean-field  method. 
We present numerical results of the single-orbital and multi-orbital models in Section \ref{numres}, by comparing the two 
magnetic phases and by examining the non-magnetic phase in detail. We summarize our results in Section \ref{conclusion}.

\section{\label{methods}Methods}


The basis of our multi-orbital tight-binding model  is spanned by 
 \(p_z\)-, \(d_{xz}\)- and \(d_{yz}\)-
orbitals localized at the sites of the honeycomb lattice.  
The other three \(d\)-orbitals couple only weakly to the \(\pi\)-bands via  \(d\)-orbital spin-orbit coupling (SOC) and 
thus can be neglected in the neighborhood of the Fermi level. 
Hopping among nearest neighboring  sites \(\langle i j\rangle \) is enabled by a term
\begin{align}
  H_0 & = \sum_{\langle i,j \rangle,\alpha,\beta,\sigma} t_{ij}^{\alpha\beta}\hat{c}^\dagger_{i\alpha\sigma} \hat{c}_{j\beta\sigma}, 
  \label{hopping_hamiltonian_real_space}
  \end{align}
where the indices \(\alpha,\beta \in \{p_z,d_{xz},d_{yz}\}\)  label the different orbitals and 
\(\hat{c}^{(\dagger)}_{i\alpha\sigma}\) are 
the annihilation (creation) operators for spin \(\sigma\). 
The hopping matrix elements are given within  the Slater-Koster approximation \cite{SlaterKoster1954},  
and the corresponding parameters are presented in Table \ref{Tight_binding_parameters}. 
\begin{table*}[!hbt]
  \caption{\label{Tight_binding_parameters}The tight-binding parameters for graphene as used in the Slater-Koster approximation \cite{GosalbezMartines2011,Konschuh2010}.}
  \begin{ruledtabular}
  \begin{tabular}{ccccccccc}
    $V_{pp\pi}$&$V_{pd\pi}$&$V_{dd\pi}$&$V_{dd\delta}$&$V_{pp\sigma}$&$V_{pd\sigma}$&$V_{dd\sigma}$&$E_d$&$E_p$\\
    \hline
  $-3$eV&$-0.69$eV&$-0.3$eV&$2.25$eV&$-8.1$eV&$3.6$eV&$3$eV&$12$eV&$0$eV\\
    \end{tabular}
    \end{ruledtabular}
    \end{table*}

Owing to symmetry, the intrinsic SOC is only non-zero for the  $d$ orbitals that constitute the $\pi$ band: 
\begin{equation}
    H^i_\text{SOC}=\xi_d \sum_{\alpha,\beta,\sigma,\sigma'}\langle \vec{L} \cdot \vec{S} \rangle_{\alpha \beta}^i \hat{c}^\dagger_{i\alpha\sigma} \hat{c}_{i\beta\sigma'}
    \label{soc_operator_matrix}
\end{equation}
with \(\xi_d = 0.8\) meV \cite{Konschuh2010}  
and $\langle \vec{L} \cdot \vec{S} \rangle_{d_{\alpha z} d_{\beta z}}^i = i\epsilon_{\alpha\beta z} s_z$, with $s_z$ being the $z$-
component of the spin. 
The on-site energy of the different orbitals is given by: 
\begin{align}
  H^i_E & = \sum_{\alpha,\sigma}E_\alpha\hat{c}^\dagger_{i\alpha\sigma} \hat{c}_{i\alpha\sigma},
\end{align}
where we choose for convenience \(E_p=0\), yielding \(E_d=12\)eV. 

For the Coulomb interaction, it is  convenient to express the density operators 
\(\hat{n}_{i\alpha\sigma} 
= \langle \hat{n}_{i\alpha\sigma} \rangle + \Delta \hat{n}_{i\alpha\sigma},
\) 
where  \(\langle \hat{n}_{i\alpha\sigma} \rangle \) is the mean value of the density and 
\(\Delta \hat{n}_{i\alpha\sigma}\) its fluctuation.  
As it is customary in the mean-field approximation,  quadratic-in-fluctuations terms are neglected, 
yielding for the  interacting Hamiltonian: 
\begin{widetext}
  \begin{align}
    H_\text{ee} &= \sum_{\sigma}\frac{U}{2}\left( \langle \hat{n}_{p\sigma} \rangle \hat{n}_{p\bar{\sigma}} + \langle \hat{n}_{p\sigma} \rangle \hat{n}_{p\bar{\sigma}} - \langle \hat{n}_{p\bar{\sigma}}\rangle \langle \hat{n}_{p\sigma} \rangle \right)\nonumber\\
    & + \sum_{n,\sigma} \Big[ V \Big(\langle \hat{n}_{p\sigma} \rangle \hat{n}_{n\bar{\sigma}} + \langle \hat{n}_{n\sigma} \rangle \hat{n}_{p\bar{\sigma}} - \langle \hat{n}_{p\bar{\sigma}}\rangle \langle \hat{n}_{n\sigma} \rangle \Big) + J\Big( \langle \hat{c}^\dagger_{n\bar{\sigma}}\hat{c}_{p\bar{\sigma}}\rangle \hat{c}^\dagger_{p\sigma}\hat{c}_{n\sigma} + \langle \hat{c}^\dagger_{p\sigma}\hat{c}_{n\sigma}\rangle\hat{c}^\dagger_{n\bar{\sigma}}\hat{c}_{p\bar{\sigma}} - \langle \hat{c}^\dagger_{p\sigma}\hat{c}_{n\sigma}\rangle \langle \hat{c}^\dagger_{n\bar{\sigma}}\hat{c}_{p\bar{\sigma}}\rangle\Big) \nonumber \\
    & + \left(V - J\right)\Big(\langle \hat{n}_{p\sigma}\rangle \hat{n}_{n\sigma} + \langle \hat{n}_{n\sigma}\rangle \hat{n}_{p\sigma} - \langle \hat{n}_{n\sigma}\rangle \langle \hat{n}_{p\sigma} \rangle - \langle \hat{c}^\dagger_{p\sigma} \hat{c}_{n\sigma} \rangle \hat{c}^\dagger_{n\sigma}\hat{c}_{p\sigma} - \langle \hat{c}^\dagger_{n\sigma}\hat{c}_{p\sigma}\rangle \hat{c}^\dagger_{p\sigma} \hat{c}_{n\sigma} + \langle \hat{c}^\dagger_{n\sigma}\hat{c}_{p\sigma}\rangle \langle\hat{c}^\dagger_{p\sigma} \hat{c}_{n\sigma} \rangle\Big)\Big], 
    \label{mean_field_full}
\end{align}
\end{widetext}
where \(p\) labels a state in a \(p_z\)-orbital and \(n\in\{d_{xz},d_{yz}\}\) labels a \(d\)-orbital. 
Here, \(U\) is the Hubbard term, that is, the on-site Coulomb repulsion for the \(p_z\)-orbitals, whereas 
\(V\) denotes the Coulomb repulsion and exchange, respectively,  
between  a \(p_z\)- and a \(d\)-orbital. 
Whereas \(U\) is on the order of \(V_{pp\pi}\), which has been obtained by comparing single-orbital tight-binding 
models with DFT calculations \cite{Pisani2007,Kuroda1987}, 
\(V\) and \(J\) 
serve here as parameters to quantify their influence on the characteristics and internal 
energy of the different phases, as it can not be obtained from experiments \cite{Yazyev2010}. 

Combining all terms described above, the total Hamiltonian for this multi-orbital tight-binding model reads
\begin{align}
  H & = H_0 + \sum_i (H^i_E + H^i_\text{SOC} + H^i_\text{ee}),
  \label{total_Hamiltonian}
\end{align}
where the first term is off-diagonal, while the rest contain on-site terms only.  
It is thus advantageous to Fourier-transform Eq. (\ref{total_Hamiltonian}) 
and solve the eigenvalue problem self-consistently via 
direct diagonalization. 
The density of states is then  numerically obtained  by sampling \(801\) points across the Brillouin zone, 
where a Gaussian broadening is optimized after a convergence study. 
A temperature of \(k_BT=3.45 \times 10^{-4}\)eV is defined for convergence purposes, allowing the computation of 
the Fermi energy \(E_F\) \cite{Claveau2014}.

\section{\label{numres}Numerical Results}
\subsection{Three different Phases}
\begin{figure*}[!hbt]
\includegraphics[angle=0,width=0.99\linewidth]{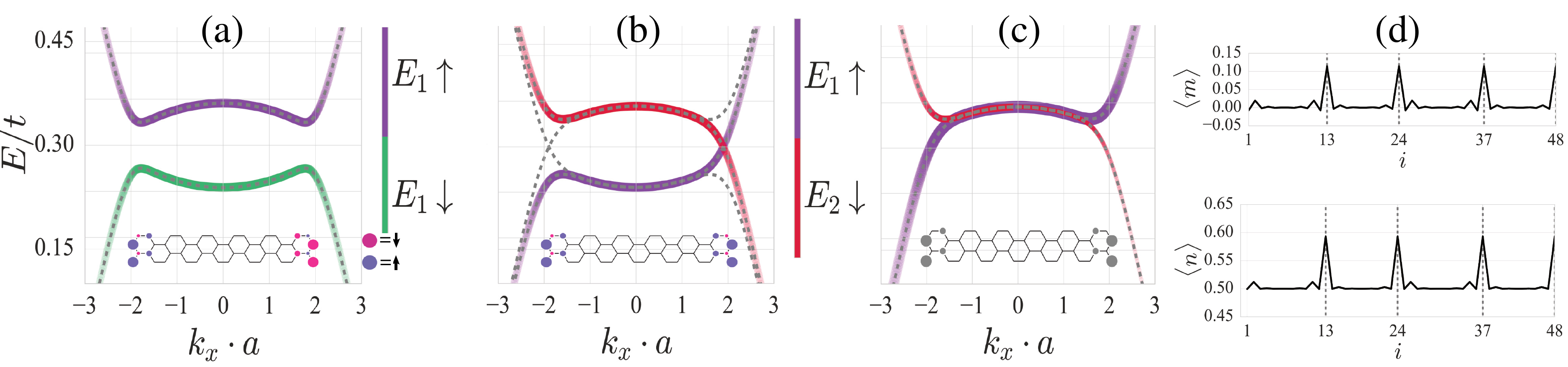}\\
\caption{
Dispersion relation of a Kramers pair of the different solutions to the mean-field problem for a \(N=12\) 
ribbon computed for the single-orbital model. The bands are colored by edge (\(E_1/E_2\)) and real spin (\(\uparrow/\downarrow\)). 
The Kramers pairs can be obtained by TR symmetry. a) AFM phase b) FM phase c) The PHS broken non-magnetic phase with \(\Delta N = 1.92\). 
Lower inset in (a, b) depict the corresponding local magnetization  or local additional occupation of (c). 
(d) Local magnetization $\langle m \rangle$ for the ground state of (a) and local density $\langle n \rangle$ corresponding to (c). 
}
\label{edge_spin}
\end{figure*}
As it is habitual in self-consistent methods, the character of the converged solution reflects that of the initial guess. 
Breaking time-reversal (TR) symmetry by starting with  parallel spin edges  leads to a magnetic phase with a finite total 
spin-magnetic moment, whereas an initial state with polarized edges of opposite sign leads to a solution with broken 
spin rotations and zero total spin. 
In both cases, half-filling is imposed. 
Another possibility (although commonly ignored), is  to consider TR-symmetric solutions 
with broken electron-hole symmetry, namely, a doped solution. 
This can be achieved by an initial state with an excess of holes or electrons, and usually converges to a
non-magnetic phase. 
In what follows, we distinguish between magnetic and non-magnetic solutions. The former occurs at half-filling 
$\langle n_i\rangle = 0.5$ whereas 
the latter has no local magnetization. The order parameter of this non-magnetic, PHS broken solution is 
given in terms of the local doping $\delta n_i$, giving  the local occupation per spin of 
$\langle n_i \rangle = 0.5 + \delta n_i $, whereas the magnetic phase is characterized 
by a finite local magnetization $\langle m_i\rangle$.  

We consider first a single-orbital ($p_z$) model in a nanoribbon with  \(N=12\) rows and \(4\) columns, using  \(U=0.6t\).   
States with local ferrimagnetic spin polarization close to the edges  
have already been theoretically obtained \cite{Lado2015,LadoPRL14, LadoPRB14,Yazyev2011PRB,katsnelson2008,Wan2011}, where
the edges may exhibit parallel (FM) or antiparallel (AFM) spin alignment with respect to each other. 
Fig. \ref{edge_spin} contains the dispersion relation of the mid-bands, where bright color indicates 
edge-localized state. 
These bands are mainly located at the edges (E1, E2) of the sample, such that TR 
partners are localized at the opposite edge \cite{Wan2011}.  
The  AFM solution in (a) corresponds to a trivial insulator, with a gap  
arising from the broken rotational symmetry \cite{Son2006b}: From a symmetry perspective, 
the \(E_2 \uparrow\) state is degenerate with the \(E_1 \downarrow\) one and vice versa, but these 
states are not related by a rotation.  From an interaction picture, 
the AFM exchange interaction acts as a staggered sublattice potential, which introduces a 
spin-dependent gap  for states localized at different edges \cite{KaneMele2005b,Ganguly2017}. 
Fig. \ref{edge_spin} (b) shows the FM solution, where two states with opposite spin cross the band gap. 
The coupling mechanism of states localized at two opposing edges can be described in analogy 
to the super exchange mechanism \cite{Jung2009b,Kunstmann2011}. 
From a topological standpoint, chiral symmetry is the reason  for the occurrence of gap-crossing states \cite{Asboth2016} 
of opposite Fermi velocity  \cite{Veyrat2020}, 
whereas from a group symmetry perspective, these solutions maintain invariance with respect to the two-fold axis, 
unlike the AFM case. 

An additional phase is encountered in this work, namely, a non-magnetic   PHS-broken solution, see 
Fig. \ref{edge_spin} (c), where the additional \(p_z\)-orbital population shifts the Fermi energy above \(U/2\). 
The electronic occupation is symmetrically distributed along the edges of the sample, 
maintaining  the original  \(C_{2\nu}\) point group of the lattice. 
This  results in a topologically non-trivial phase, just like in the non-interacting spin-Hall insulating case \cite{KaneMele2005a}. 
The corresponding order parameters are depicted  in Fig. \ref{edge_spin} (d), that is, the 
on-site magnetization 
 \(\langle m_i \rangle = \frac{1}{2}(\langle \hat{n}_{i,\uparrow} \rangle - \langle \hat{n}_{i,\downarrow} \rangle)\) 
for the magnetic phases and the occupation for the PHS-broken solution, 
\(\langle n_i \rangle = \frac{1}{2}(\langle \hat{n}_{i,\uparrow} \rangle + \langle \hat{n}_{i,\downarrow} \rangle)\). 
Values are plotted as a function of the site index \(i\)  in sequence that follows columns (left to right) and rows (top to bottom) of 
the unit cell depicted in the insets of Fig. \ref{edge_spin} (a-c). 
For the FM solution, the magnetization is largest at the edge sites, with an exponential decay towards the bulk. 
The AFM phase (not shown) has similar magnetization, only with opposite sign at each edge. 
The PHS broken solution lacks of magnetic moments, however,  
the electronic occupation off half filling shows a similar behavior as  \(\langle m\rangle \)
in the magnetic solutions. We stress that a similar solution with negative occupation off half filling (or 
hole-doped) is also found when the initial state has a filling slightly below half filling. 

It is worth mentioning that the topologically trivial AFM solution minimizes the free energy 
when the half-filling condition is imposed. However, dopants change this criterion and allows for a PHS broken state. 
In what follows we examine the free energy, the local magnetization 
and the occupation of these three solutions as a function of ribbon size and interaction parameters in single and  
multi-orbital approach.

\subsection{Energetic considerations}
Lieb's theorem states that the ground state at half-filling has zero spin-magnetic moment \cite{Lieb1989}. 
It follows that the ground state at half-filling is always the AFM edge phase, 
regardless of ribbon size and for all \(U>0\). 
However, Lieb's theorem does not apply for doped systems, allowing for 
a ground state without any local spin-magnetic moment. 
We first compute and analyze the free energy of the three phases within the single-orbital model, and extend 
our computation  to  the multi-orbital model. 
\subsubsection{ Single-orbital model}
\begin{figure}[!hbt]
    \includegraphics[width=.99\linewidth]{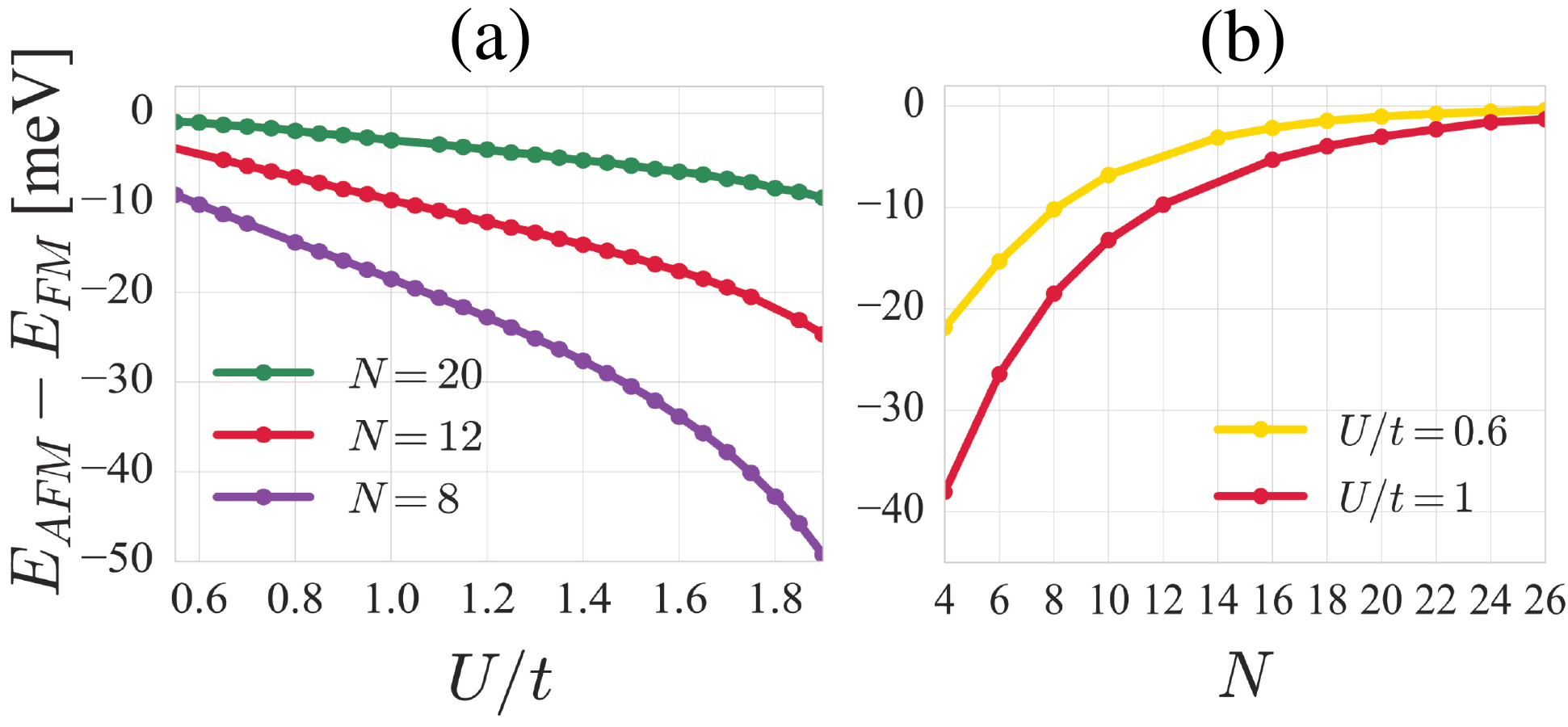}\\
\caption{Energy differences of FM and AFM phases for (a) \(N=8\),  \(12\) and \(20\) as a function of  \(U\) and 
(b) for \(U/t=0.6\) and \(U/t=1\), as a function of  number of rows \(N\).} \label{energy_difference} 
\end{figure}

We first employ a single-orbital ($p_z$) model, where $t_{ij}  = V_{pp\pi}\equiv t$ in Eq. (1),  and 
with the Hubbard term being the only interacting term. 
We compute the free energy of the ground state for the different solutions. 
In Fig. \ref{energy_difference} (a) the energy differences of the two magnetic phases at 
half-filling for three different ribbon sizes \(N=8\), \(12\) and \(20\) is shown as a function of  \(U\), 
where the AFM phase is found to always be lower in energy. 
As expected, the energy separation between both phases increases with the interaction parameter $U$.  
For \(N=8\), the reduced edge separation results in a relative lower energy of the AFM  
due to the inter-edge super exchange mechanism \cite{Jung2009b}. 
For larger \(N\) this energy gain becomes smaller and a law \(\sim N^{-2}\) can be seen in 
Fig. \ref{energy_difference} (b) for both \(U/t = 0.6\) and \(U/t = 1\).
Hence, it is to be expected that in the thermodynamic limit $N\to\infty$  
both magnetic phases become equally probable, as $E_{\rm FM} - E_{\rm FM} \ll k_BT$. 

\begin{figure}[!hbt]
    \includegraphics[width=.6\linewidth]{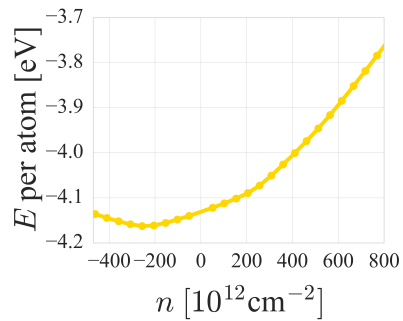}\\
\caption{The energy of the PHS broken phase for different amounts of doping for \(N=8\) and \(U/t=0.6\) in the single-orbital model.} \label{nonmag} 
\end{figure}

We consider next the PHS broken solutions, where the Stoner criterion justifies the absence of magnetism in doped systems, 
as  an increase of the Fermi energy above the peak in the density of states 
removes the magnetic instability \cite{Fazekas1999, FernandesRossier2008}. 
We examine the behavior of  the energy of the non-magnetic phase 
as the filling is shifted by an amount $\delta n$ off half-filling. 
Fig. \ref{nonmag} shows the free energy as a function of doping for a \(N=8\) and \(U=0.6\) nanoribbon.  
Here, the doping is expressed in cm$^{-2}$ units, $n = \delta n/S$  where \(S=6.24 \times 10^{-15}\)cm\(^2\) is the area of the ribbon.    
For low $\delta n$, the free energy obtained numerically has a quadratic dependency on the doping,  
\begin{align}
  F(U,n) & = UN \left(1 +\frac{ nS}{4N}\right)^2,  
  \label{energy_shift_doping}
\end{align}
which corresponds to the interacting term of Eq. (\ref{mean_field_full}). 
For larger doping, \(|n| > 250\) \(\times 10^{12}\)cm\(^{-2}\), a
stronger quadratic dependency on $n$ is observable. Here, the kinetic energy gain  given by $H_0$ is larger than the 
effective reduction via Coulomb interaction, giving rise to a strong quadratic increase in energy. 
This results in a minimum for the free energy, which is encountered in the hole doping regime at the boundary between small and large filling. 

\begin{figure}
    \includegraphics[width=\linewidth]{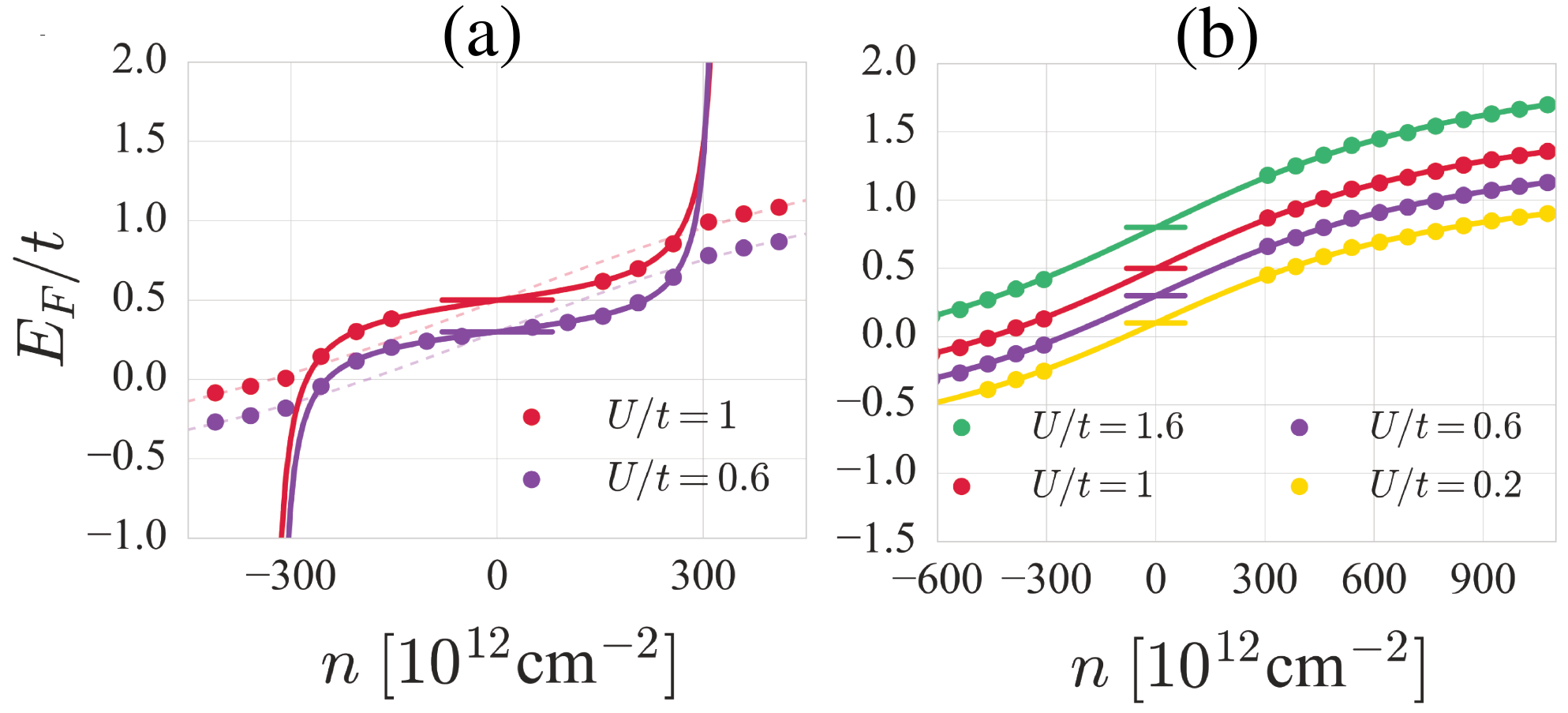}
\caption{
The Fermi energy of the PHS broken solution for \(N=8\) and different \(U/t\). 
The horizontal bar indicates the Fermi energy of the two magnetic phases at half-filling. 
a) For smaller doping the Fermi energy is fitted via \(\tan{n}\). 
The dashed lines correspond to a fit via \(\arctan{n}\). b) Fit with \(\arctan{n}\) for larger doping, 
for four different Hubbard parameters: $U/T =$ 0.2, 0.6, 1.0 and 1.6.} \label{doping} 
\end{figure}
Fig. \ref{doping} (a) shows the Fermi energy as a function of  \(U/t\), where the two doping regimes are visible. 
The separation between the small and large doping regimes occurs around \(|n| \approx 250\) \(\times 10^{12}\)cm\(^{-2}\). 
A function \(\tan{n}\) fits very well to the data of the small doping regime, which is indicated in the figure by solid lines. 

For larger doping, however, the Fermi energy has a different dependency, namely \(\arctan{n}\), given by dashed lines in Fig. \ref{doping} (a). 
This dependency is still visible when \(U/t\) is varied between \(0.2\) and \(1.6\) for dopings up to 
\(n =  10^{15}\)cm\(^{-2}\), as shown in Fig. \ref{doping} (b). 
While these results have been obtained in small systems, this behavior should scale to larger ribbons. 
The amount of doping at which the bulk states become filled 
in larger nanoribbons is reduced, as the band separation decreases with a power law \cite{Wakabayashi2010} of 
\(\sim 1/N\) . 
This suggests that in larger graphene nanostructures, the doping  required to reach the energetic minimum decreases, which may serve as 
an explanation to the
elusive measurement of magnetic edges in graphene ribbons, 
which was previously unaccounted for. 

\subsubsection{Multi-orbital model}
We consider next the multi-orbital model 
by including the \(d_{xz}\)- and \(d_{yz}\)-orbitals, which breaks PHS.  
We define $\delta n_p$ and $\delta n_d$ as the \(p_z \)- and \(d \)-orbital doping, respectively. 
The half-filling condition reads $\delta n_p= -\delta n_d$, 
where the orbital occupation per spin is given by 
\( n_p =  0.5 + \delta n_p \) and 
\( n_d =  \delta n_d \), respectively. 
This leads to a renormalization of the orbital energies, namely: 
\begin{align}
  E_\alpha(\delta n_\alpha) & = \sum_{\sigma}\left[E_d + \delta n_\alpha\left(2V - J\right)\right] \hat{n}_{\alpha\sigma}, 
  \label{d_energy_shift_half_filling}
\end{align}
where we can infer that a decreased energetic separation of $p_z$ and $d$-orbitals lead to enhancement of the d-band population 
and vice-versa. 
\begin{figure}[!hbt]
    \includegraphics[width=\linewidth]{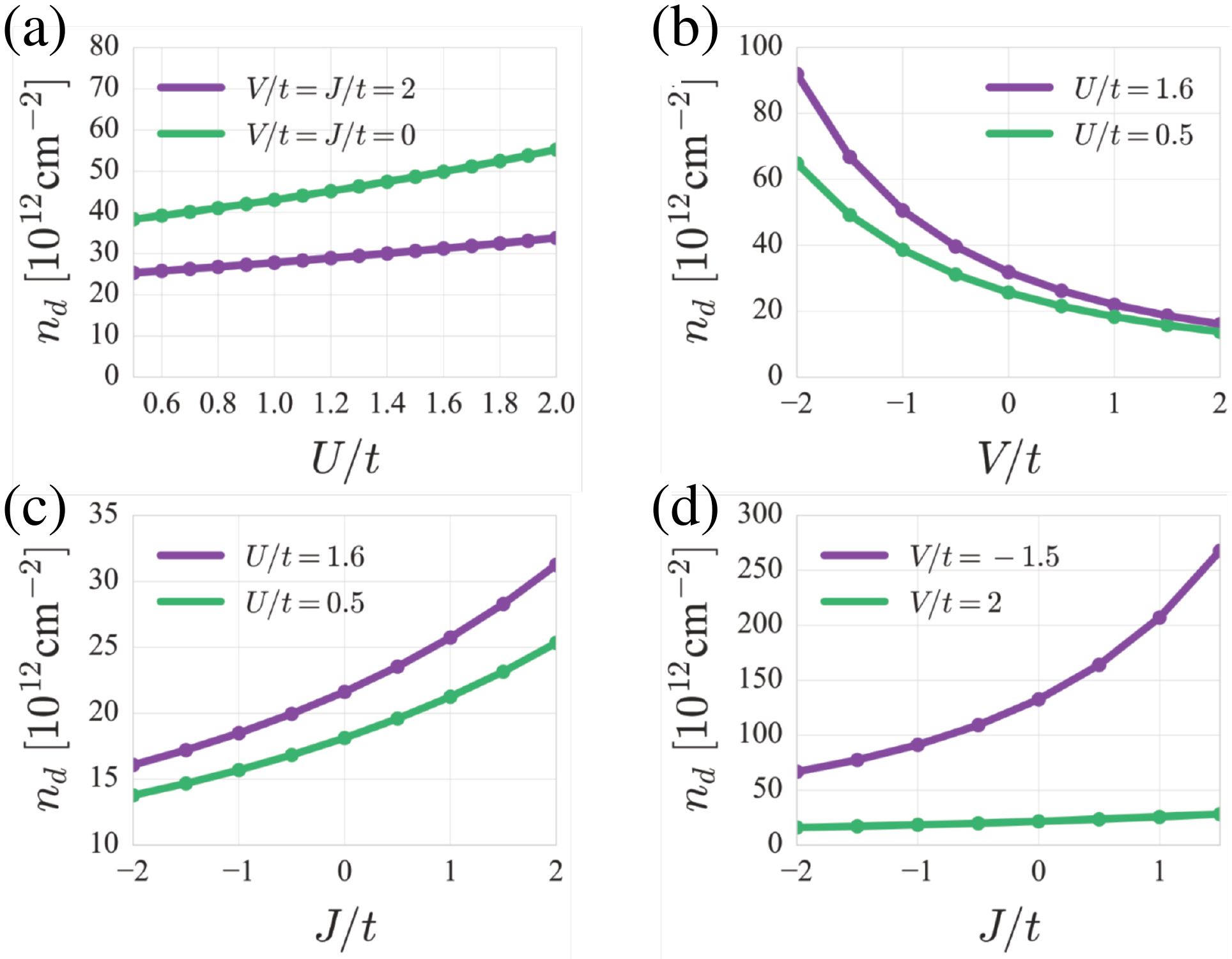}
\caption{Total \(d\)-state occupation of a ribbon with \(N=8\) plotted for different interacting parameters. 
Both FM and AFM phase have identical occupation.} \label{d_occupation} 
\end{figure}

Fig. \ref{d_occupation} shows $n_d$ as a function of the interacting parameters in units of $t$. 
When \(V\) and \(J\) are kept constant, an increased \(U\) leads to a larger occupation of the \(d\)-orbitals 
[see Fig.  \ref{d_occupation} (a)], 
as the Coulomb repulsive interaction is larger at  the occupied $p_z$ orbitals.   
For negative $V$ or positive $J$, the \(d\)-orbital occupation becomes larger, as the \(d\)-orbital energy shift is reduced, as is shown in (b) 
and (c), respectively. 
Finally, Fig. \ref{d_occupation} (d) shows that typical positive $V$ values yield small \(d\)-band occupation for any $J$, whereas 
negative $V$ results in $d$ population up to two orders of magnitude larger. We stress, however, that 
the repulsive (physical) Coulomb interaction yields $V, J>0$. 

\begin{figure}[!hbt]
  \includegraphics[width=.7\linewidth]{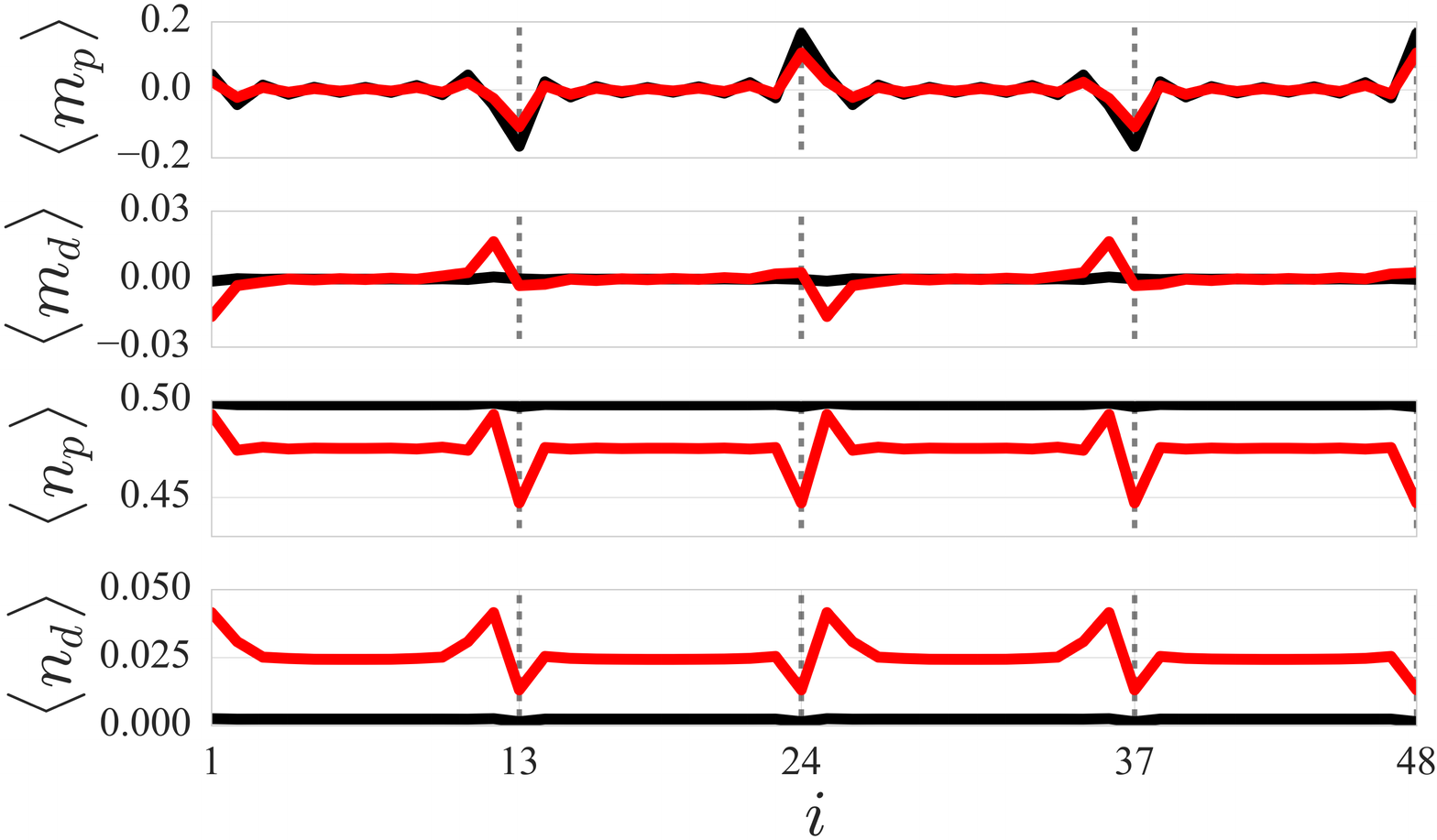}
  \caption{The expectation value of the magnetization \(m_{p/d}\) and occupation \(n_{p/d}\) per site and spin for a ribbon 
  with \(N=12\) in the AFM phase, where \(U=1.6t\) and \(J=0.5t\), with \(V=-2t\) (red) and \(V=2t\) (black).}
\label{magnetization_d}
\end{figure}

 Fig. \ref{magnetization_d} shows the local magnetization and orbital doping of a 48-sites ribbon in the  AFM phase, with
\(U=1.6t\) and \(J=0.5t\). An attractive $d$-orbital Coulomb interaction $V=-2t$ (red) results in an enhanced $d$-orbital 
magnetization at the sites next to the edge, whereas the single orbital results are recovered for 
a repulsive $d$-orbital Coulomb interaction $V=2t$ (black).

\begin{figure}
  \includegraphics[width=\linewidth]{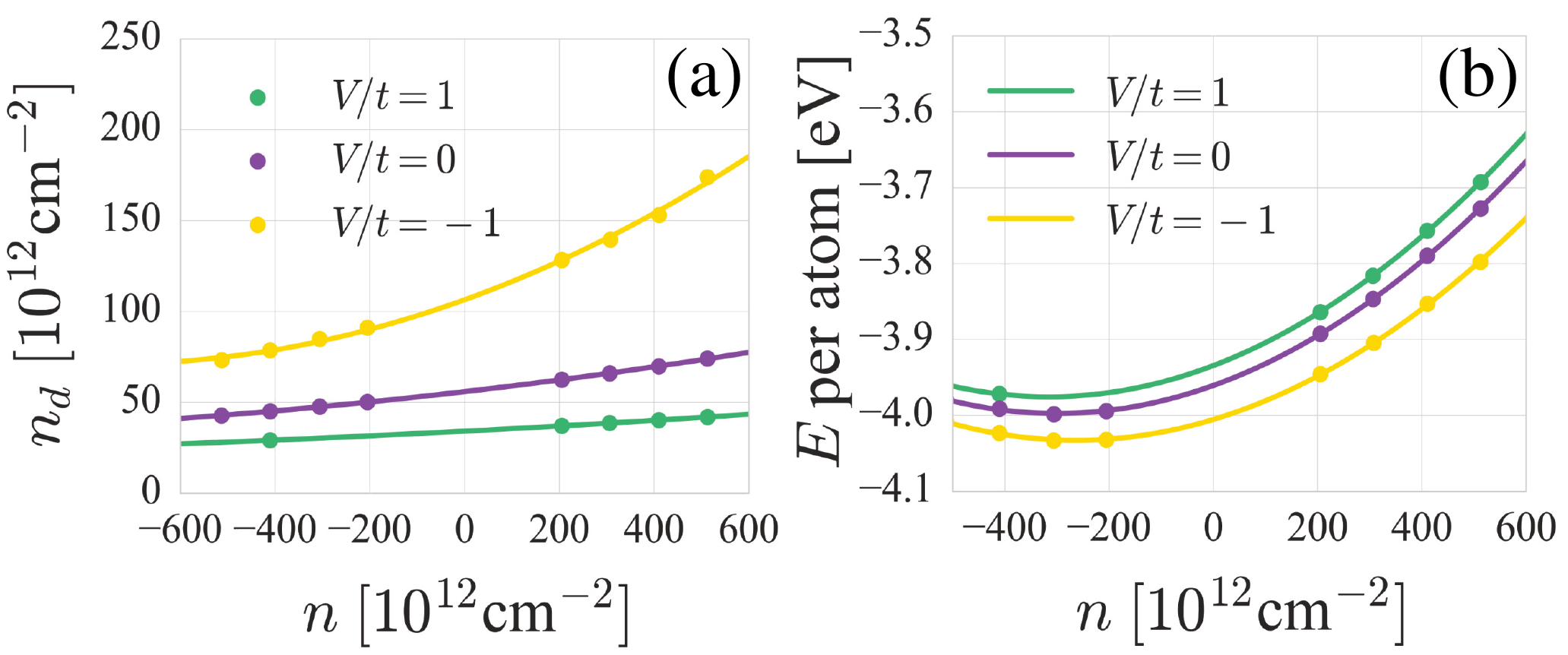}
  \caption{Results for a nanoribbon with \(N=8\), for \(U/t=1\) and \(J/t=1\). a) Total \(d\)-state occupation and b) energy of the non-magnetic phase for different doping strengths \(n\) for \(V/t = 1\), \(0\) and \(-1\).}
\label{doping_d}
\end{figure}
Finally, we consider the multi-orbital doped solution. 
Fig. \ref{doping_d} (a) shows the $d$-orbital doping as a function of total doping, $n = (\delta n_p + \delta n_d)/S$ for 
positive (green), zero (purple) and negative (yellow) $p$-$d$ Coulomb interaction, $V$. A quadratic law is observed for the latter, 
whereas a slow, linear increase in $d$-orbital occupation occurs for the former. 
The free-energy functional at low doping 
results now in: 
\begin{align}
  &F(n,U,V,J) = UN \left(1 +\frac{ \delta n_p }{4N}\right)^2   + \left(2 V - J \right) n_p\delta n_d, 
  \label{energy_doping}
\end{align}
where $n_p = nS -\delta n_d$.  
The first term dominates, and was already observed in the single-orbital model [see Eq. (\ref{energy_shift_doping})]. 
For larger dopings, the kinetic energy raises, and the single-orbital results are recovered, with the total minimum appearing for 
similar doping values. 
The inclusion of \(d\)-orbitals does not lead to qualitative differences for the free energy or the character of the ground state, compared to the single-orbital model. However, regarding the additional orbitals is imperative in low-energy considerations, such as the size of the intrinsic SOC band gap, which is linear with the 
$d$-orbital occupation \cite{Konschuh2010}. 

\section{\label{conclusion}Conclusion}
We have examined three different edge phases in terms of local magnetization and local orbital population, 
both in the single and in the multi-orbital model, of graphene nanoribbons with zigzag edges. 
For half-filling and finite interaction strengths the inter-edge super exchange mechanism predicts a lower free energy of the AFM phase compared to the FM phase. 
Our mean-field numerical calculations on graphene nanoribbons confirm these results. 
The PSH symmetry broken phase shows a minimum in the free energy at  the boundary between the small and  the large 
hole doping regime.
In the former regime, the local doping occurs exclusively at the edge atoms, lowering the energy via exchange, whereas in the latter, the doping spreads throughout the bulk, 
lowering the kinetic energy and hence, increasing the free energy.  
These result are unchanged in a multi-orbital model and we have also briefly discussed the expected validity of our findings for larger systems.
Thus, the multi-orbital computations confirm that the single-orbital description remains a good approximation for the energy gap of the two magnetic phases, whereas the \(d\)-orbital occupation only becomes relevant at attractive Coulomb interaction parameters.
In the thermodynamic limit (large $N$, small interactions) the free energy of the three edge phases coincide, and hence, observation of a magnetic state would require very low temperatures, small ribbon sizes and specific low-interacting, acceptor substrates. 

\section{\label{acknowledgements}Acknowledgements}
This work is supported/funded by the Cluster of Excellence 
"CUI: Advanced Imaging of Matter" of the Deutsche Forschungsgemeinschaft (DFG) – EXC 2056 – project ID 390715994.
We thank A. Chudnovskiy and R. H. Blick for fruitful discussions.

\bibliography{bib_full_hf}

\end{document}